\documentclass[twocolumn,showpacs,preprintnumbers]{revtex4}%
\usepackage{amsfonts}
\usepackage{amsmath,epsfig}
\usepackage{graphicx}
\usepackage{dcolumn}
\usepackage{bm}
\usepackage{amssymb}
\usepackage{amsmath}%
\begin{document}
\title{Is there a ``$\Delta$-isobar puzzle'' in the physics of neutron stars?}
\author{Alessandro Drago$^{\text{(a)}}$, Andrea Lavagno$^{\text{(b)}}$, Giuseppe Pagliara$^{\text{(a)}}$ and Daniele Pigato$^{\text{(b)}}$}
\affiliation{$^{\text{(a)}}$Dip.~di Fisica e Scienze della Terra dell'Universit\`a di Ferrara \\
INFN Sez.~di Ferrara, Via Saragat 1, I-44100 Ferrara, Italy}
\affiliation{$^{\text{(b)}}$ Department of Applied Science and Technology, Politecnico
di Torino, Italy \\
Istituto Nazionale di Fisica Nucleare (INFN), Sezione di Torino, Italy}

\begin{abstract}
We discuss the formation of $\Delta$ isobars in neutron star
matter. We show that their threshold density strictly correlates with
the density derivative of the symmetry energy of nuclear matter, the
$L$ parameter. By restricting $L$ to the range of values indicated by
recent experimental and theoretical analysis, i.e. $40$ MeV $\lesssim
L \lesssim 62$ MeV, we find that $\Delta$ isobars appear at a density
of the order of 2$\div$3 times nuclear matter saturation density,
i.e. the same range for the appearance of hyperons. The range of
values of the couplings of the $\Delta$s with the mesons is restricted
by the analysis of the data obtained from photoabsorption, electron
and pion scattering on nuclei. If the potential of the $\Delta$
in nuclear matter is close to the one indicated by the experimental
data then the equation of state becomes soft enough that a ``$\Delta$
puzzle'' exists, similar to the "hyperon puzzle'' widely discussed in
the literature.
\end{abstract}

\maketitle

Since the seminal paper of Ref.~\cite{Glendenning:1984jr}, the possible
formation in the core of neutron stars of baryons heavier than the
nucleon is one of the most interesting open issue in nuclear astrophysics.
While a huge literature is available concerning the appearance of
hyperons in neutron stars (see for instance Ref. \cite{Weissenborn:2011ut}
and references therein) only little work has been done to asses whether
$\Delta$(1232) isobars can also take place in those stellar objects
\cite{Huber:1997mg,Xiang:2003qz,Chen:2007kxa,Chen:2009am,Schurhoff:2010ph,Drago:2013fsa,Pagliara:2014gja}.
The reason why $\Delta$ resonances have been neglected is maybe connected with the
outcome of Ref.~\cite{Glendenning:1984jr} indicating
that these particles would appear at densities much higher than
the typical densities of the core of neutron stars and they are
therefore irrelevant for astrophysics. On the other hand, hyperons
could appear already at 2$\div$3 times the density of nuclear matter,
$n_0=0.16$ fm$^{-3}$, and one has to include these degrees of freedom
when modeling the equation of state of dense nuclear matter.
The consequent softening of the equation of state reduces the maximum
mass of neutron stars which in many calculations drops
below the $2M_{\odot}$ limit imposed by the
precise measurements of the masses of PSR J1614-2230 and PSR
J0348+0432 \cite{Demorest:2010bx,Antoniadis:2013pzd}.
This inconsistency between
astrophysics (mass measurements) and hadron physics (the necessary
appearance of new degrees of freedom at large densities) is re-known as
the "hyperon puzzle''. However the uncertainties on the hyperons'
interactions in dense matter are such that it is still possible to
tune the parameters, within phenomenological models, in order to
fulfill the $2M_{\odot}$ limit also when hyperons are included in the
equation of state
\cite{Weissenborn:2011ut,Weissenborn:2011kb,Bednarek:2011gd,vanDalen:2014mqa}. On the
other hand in microscopic models based on the Brueckner-Hartree-Fock
approach, even three body forces are not enough to allow for the existence of
massive stars \cite{Vidana:2010ip} although more sophisticated calculations
based e.g. on Monte Carlo techniques are needed before a firm conclusion
can be drawn \cite{Lonardoni:2013gta}.

In principle, also the appearance of $\Delta$ isobars at some critical
density $n_{\rm crit}^{\Delta}$, softens the equation of state thus reducing the maximum mass
with respect to the case in which those particles are simply
neglected. The crucial question, that we investigate in this paper,
concerns the value of $n_{\rm crit}^{\Delta}$ in beta-stable matter. We will show, in
particular, that a significant correlation exists between $n_{\rm crit}^{\Delta}$ and the
density derivative of the symmetry energy $S$, the
parameter $L=3 n_0 (\mathrm{d}S)/(\mathrm{d}n_B)$.
It will be clear in the following that
the appearance of the $\Delta$ isobars is affected by the value
of $S$ at a density close to $n_{\rm crit}^{\Delta}$. Since the value of $S$
at $n_0$ is determined with a good precision, the crucial quantity
becomes $L$. Only recently it has been possible to strongly constrain
the value of $L$ both from terrestrial and astrophysical data
\cite{Lattimer:2012xj} with the result that $40.5\lesssim L \lesssim
61.9$ MeV.

First we will show, within a toy model equation of state
based on the GM3 model of Ref.\cite{Glendenning:1991es}, the existence
of a correlation between $L$ and $n_{\rm crit}^{\Delta}$; then, by using a ``state
of the art'' equation of state built upon several experimental nuclear
physics information \cite{Steiner:2012rk}, we will calculate
$n_{\rm crit}^{\Delta}$ and show that $\Delta$ isobars appear actually before the
hyperons and they should be included in the dense matter equation of
state. We will also discuss the experimental constraints on the $\Delta$
couplings in nuclear matter obtained from photoabsorption, electron and pion scattering
data and finally we will compute the effect of the appearance of these
additional baryons on the maximum mass and on the radii of compact stars.

We adopt here the scheme of relativistic mean model in which the
interaction between baryons is mediated by the exchange of a scalar
meson $\sigma$, an isoscalar vector meson $\omega$ and a isovector
vector $\rho$. The threshold for the formation of the i-th baryon is given
by the following relation:
\begin{equation}
\mu_i \ge m_i-g_{\sigma i} \sigma + g_{\omega i}\omega + t_{3 i}g_{\rho i} \rho
\label{soglia}
\end{equation}
where  $\sigma$, $\omega$ and $\rho$ are the expectation values of the corresponding fields,
$g_{\sigma i}$, $g_{\omega i}$, $g_{\rho i}$ are the couplings between the mesons and the baryons,
$\mu_i$, $m_i$ and $t_{3 i}$ are the chemical potential, the mass and the isospin charge of
the baryons.
The baryon chemical potential $\mu_i$ are obtained by the $\beta$-equilibrium
conditions: $\mu_i=\mu_B+c_i\,\mu_{C}$, where $\mu_B$ and $\mu_{C}$ are
the chemical potentials associated with the conservation of the baryon
number and the electric charge respectively and $c_i$ is the electric
charge of the i-th baryon.

As already extensively discussed in Ref. \cite{Glendenning:1984jr}, among the four $\Delta$ isobars, the $\Delta^-$
is likely to appear first, in $\beta$-stable matter, because it can replace a neutron and an electron at the top of their Fermi seas.
However, this particle is ``isospin unfavored'' because its
isospin charge $t_3=-3/2$ has the same sign of the isospin charge of
the neutron. For large values of the symmetry energy $S$ and, therefore, of $g_{\rho \Delta}$, the $\Delta^-$ appears
at very large densities or it does not appear at all in dense matter
thus playing no role in compact stars. Indeed, in Ref. \cite{Glendenning:1984jr} the $\Delta$ isobars could appear in neutron stars only for not physical small values of the symmetry energy,
obtained by setting $g_{\rho i}=0$ for all the baryons.

\begin{figure}[ptb]
\vskip 0cm
\begin{centering}
\epsfig{file=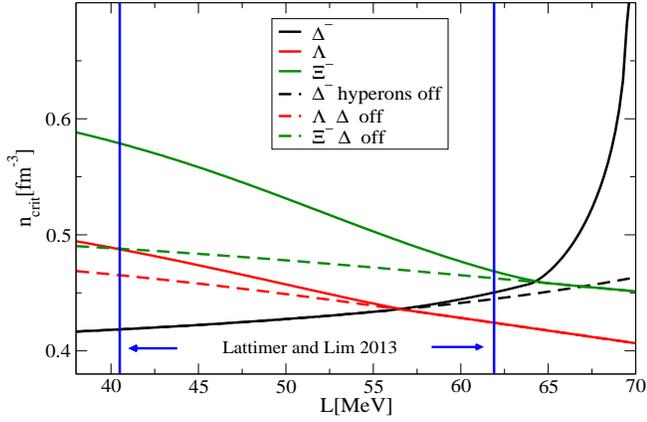,height=8.5cm,width=5.5cm,angle=-90}
\caption{(Color online) Threshold densities of hyperons and $\Delta$s as functions of the $L$ parameter. The continuous lines
refer to the case in which all the degrees of freedom are included in the computation of the equation of state and the dashed lines
refer to the case in which either hyperons or $\Delta$s are artificially switched off. The vertical lines
indicate the range of allowed values of $L$ as found in \cite{Lattimer:2012xj}.}
\label{soglie}
\end{centering}
\end{figure}

The lagrangian adopted in
\cite{Glendenning:1984jr} is a Walecka-type model with
minimal coupling terms between baryons and the $\omega$ and $\rho$
mesons and linear and non-linear interaction terms for the scalar meson
$\sigma$. In such a scheme the symmetry energy reads $S =
S_{kinetic}+S_{interaction}$, where the interaction term
$S_{interaction} = \frac{g_{\rho N}^2}{8m_{\rho}^2} n_B $,
$m_{\rho}$ is the mass of the $\rho$ meson and $n_B$ is the baryon density.
The coupling $g_{\rho N}$ (where the label $N$ stands for the nucleon) is fixed by using the experimental value of
the symmetry energy, the most recent estimates ranging in the interval
$29 \lesssim S \lesssim 32.7$ MeV \cite{Lattimer:2012xj}. In this
scheme no experimental information on the density dependence of the
symmetry energy can be incorporated and in particular the $L$ parameter is
automatically fixed once a specific value of $S$ is adopted. It turns
out, that in the models introduced in Ref.s \cite{Glendenning:1984jr,Glendenning:1991es}, $L \sim 80$ MeV and is thus
higher than the values suggested by the most recent analysis
\cite{Lattimer:2012xj}. There are two ways to modify the lagrangian
adopted in the GM models in order to include the new experimental
information: to introduce
density dependent couplings or to introduce non minimal couplings also
for the vector mesons, the two approaches being basically equivalent.

\begin{figure}[ptb]
\vskip 0cm
\begin{centering}
\epsfig{file=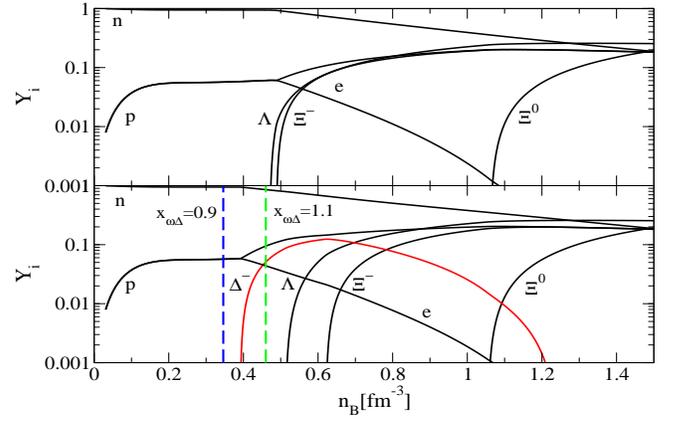,height=8.5cm,width=5.5cm,angle=-90}
\caption{(Color online) Particles fractions as functions of the baryon density: only hyperons (upper panel),
hyperons and $\Delta$s (lower panel) for $x_{\sigma\Delta}=x_{\omega\Delta}=x_{\rho\Delta}=1$. The red line indicates
the fraction of the $\Delta^-$ which among the four $\Delta$s are the first to appear.
The blue and the green vertical lines indicate the onset of the formations of $\Delta^-$ for $x_{\omega\Delta}=0.9$ and  $x_{\omega\Delta}=1.1$, respectively.}
\label{particles}
\end{centering}
\end{figure}

In our first analysis we adopt the GM3 model but we consider a density
dependent baryon-$\rho$ meson coupling i.e.  $g_{\rho i}=g_{\rho
  i}(n_0)e^{-a(n_B/n_0-1)}$ (see Ref.~\cite{Typel:2009sy}). In this
way we introduce a single parameter $a$ which affects only the value
of $L$ leaving untouched the other properties of nuclear matter at
saturation.  As customary, for the couplings of hyperons and $\Delta$
isobars with the mesons, we introduce the ratios $x_{\sigma i}=g_{\sigma\, i}/{g_{\sigma N}}$,
$x_{\omega i}=g_{\omega\, i}/{g_{\omega N}}$ and $x_{\rho i}=g_{\rho\, i}/{g_{\rho N}}$, 
where the index $i$ runs over all the hyperons and $\Delta$ isobars. For simplicity, we start by fixing these ratios
to 1 for the $\Delta$ isobars, as in Ref.~\cite{Glendenning:1984jr}.  We
will later show that $x_{\sigma\Delta} \simeq x_{\omega\Delta} \simeq 1$
are compatible with the experimental data coming from electron and
pion scattering on nuclei and photoabsorption nuclear reactions.  For the hyperons we use the same values as
in Ref.~\cite{Drago:2013fsa} obtained by reproducing the binding
energies of the hyperons in ipernuclei and by imposing the SU(6)
symmetry.

We can now study how the values of $n_{\rm crit}^B$ for the
different baryons change as a function of the new parameter $a$ or,
equivalently, as a function of $L$.  We limit our discussion to
the case of the $\Lambda$, $\Delta^-$ and  $\Xi^-$ which
are the first heavy baryons appearing as the density increases.  The
results are displayed in Fig. \ref{soglie}. One can notice the
different behavior of the
thresholds: the larger the value of $L$ the larger $n_{\rm crit}^{\Delta}$
and the smaller $n_{\rm crit}^{\Lambda}$ and $n_{\rm crit}^{\Xi^-}$.
Indeed, for growing values of $L$, the isospin term in Eq.~(\ref{soglia}) also increases and the
$\Delta$ isobar becomes more and more isospin unfavored. Even though the $\Lambda$
is not directly coupled to the $\rho$ meson ($t_{3\Lambda}=0$), the value of $L$ still affects
$n_{\rm crit}^{\Lambda}$ defined by the equation $\mu_{\Lambda}(k_{F}^{\Lambda}=0)=\mu_n(n_B=n_{\rm crit}^{\Lambda})$.
More explicitly this equation reads
 $x_{\omega\Lambda}g_{\omega
  n}\omega+m^{*}_{\Lambda} =g_{\omega
  n}\omega-\frac{1}{2}g_{\rho n}\rho+\sqrt{k_{F n}^{2}+m^{*\,2}_n}$.
The SU(6) symmetry implies $x_{\omega\Lambda}=2/3$ and the equation simplifies to:
$m^{*}_{\Lambda} =g_{\omega
  n}\omega/3-\frac{1}{2}g_{\rho n}\rho+\sqrt{k_{Fn}^2+m^{*\, 2}_n}$
where the
mean field value $\omega$ is positive being proportional to the baryon density. 
The mean field $\rho$ is proportional to the
difference between protons and neutrons and it is therefore negative.
Clearly larger values of $g_{\rho n}$ (or equivalently of $L$) imply smaller values
of $n_{\rm crit}^{\Lambda}$.
Similarly for the $\Xi^-$ the threshold equation reads: $\mu_{\Xi^-}(k_{F}^{\Xi^-}=0)=\mu_n(n_B=n_{\rm crit}^{\Xi^-})+\mu_e$.
Again by using SU(6), $x_{\omega\Xi^-}=1/3$ and $x_{\rho\Xi^-}=1$, and the
threshold reads:
$\frac{2}{3}g_{\omega n}\omega+\sqrt{k_{Fn}^2+m^{*\,2}_n}+\mu_e =
m^{*}_{\Xi^-}$.
Larger values of $L$ imply larger amounts of
protons and electrons, thus $\mu_e$ increases as a function of $L$ and the appearance of
the $\Xi^-$ is favored. Finally for the $\Delta^-$,
$\mu_{\Delta^-}(k_{F}^{\Delta^-})=\mu_n(n_B=n_{\rm crit}^{\Delta^-})+\mu_e$ and the threshold conditions (assuming
all the ratios $x_{i\Delta}=1$) reads: $g_{\rho
  n}\rho+\sqrt{k_{Fn}^2+m^{*\,2}_n}+\mu_e = m^{*}_{\Delta^-}$ and,
contrary to the case of the $\Lambda$ and $\Xi^-$, larger values of $L$ lead to larger
values of $\mu_e$ but, at the same time, also to larger values of the (negative)
quantity $g_{\rho n}\rho$. Notice that this term is twice as large but with the opposite sign of the similar term
appearing in the equation for the $\Lambda$. The $L$ dependence of $n_{\rm crit}^{\Delta^-}$ is therefore dominated by $g_{\rho n}\rho$.

The dashed lines in Fig. 1 correspond to the $n_{\rm crit}^{i}$ in the cases in which
either the hyperons or the $\Delta$ isobars are artificially excluded in
the computation of the equation of state.  In particular one can
notice, that at high values of $L$, larger than about $65$ MeV, the
threshold of the $\Delta^-$ increases very rapidly with $L$. This
corresponds to the values of $L$ for which the $\Xi^-$ appears before
the $\Delta^-$ thus completely suppressing those particles. Indeed
within the GM3 model, for which $L\sim 80$ MeV, the $\Delta^-$ do not
appear at all as already found in Ref. \cite{Glendenning:1984jr}. Similarly, one can notice that if the
isobars are formed before the hyperons, what happens at $L\sim 56$
MeV, $n_{\rm crit}^{\Lambda}$ and $n_{\rm crit}^{\Xi^-}$ are shifted to larger
densities, as already noticed in Ref. \cite{Drago:2013fsa}. Similar results
have been found in \cite{Glendenning:1984jr}, where two cases are
analyzed, corresponding to a finite and to a vanishing value of
$g_{\rho n}$, with the result that in the case of $g_{\rho n}=0$ the
isobars are favored. Finally, the blue lines mark the range of the
values of $L$ indicated by the analysis of Ref.~\cite{Lattimer:2012xj}:
the recent constraints on $L$ imply that at densities close to three
times $n_0$ both the hyperons and the isobars must be included in the
equation of state and for the lower allowed values of $L$, the isobars
appear even before the hyperons. This will have consequences both for
cold and catalyzed neutron stars, as we will show in the following,
and for protoneutron stars evolution.
Finally let us stress that all the previous analysis are based
on a rather conservative choice for the couplings between $\Delta$s and mesons.
If higher values of $x_{\sigma,\Delta}$ and or lower values for  $x_{\omega,\Delta}$ are adopted, 
$n_{\rm crit}^{\Delta}$ can result to be smaller than $n_{\rm crit}^{\Lambda}$ and $n_{\rm crit}^{\Xi^-}$ for all the reasonable values of $L$.

\begin{figure}[htb]
\vskip 0cm
\begin{centering}
\epsfig{file=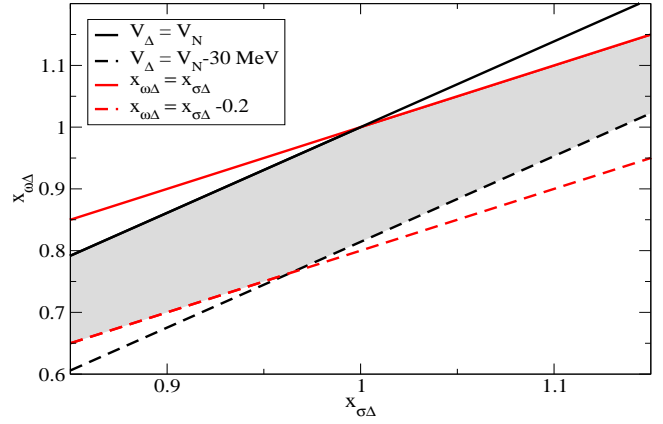,height=8.5cm,width=5.5cm,angle=-90}
\caption{(Color online) Relation between the coupling ratios $x_{\omega\Delta}$ and $x_{\sigma\Delta}$ for two
values of the potential $V_{\Delta}$ as obtained from pion and electron scattering and from photoabsorption on nuclei. 
Also the experimental constraints on the difference between $x_{\omega\Delta}$ and $x_{\sigma\Delta}$ are displayed \cite{Wehrberger:1989cd}. 
The grey area corresponds to the region in which all constraints are satisfyed.}
\label{be}
\end{centering}
\end{figure}

Let us turn now to the more sophisticated model for the equation of
state proposed in Ref. \cite{Steiner:2012rk}. In the corresponding
lagrangian, self interaction terms for the vector mesons and mixing
terms between the scalar and the vector sectors are added
\cite{Steiner:2004fi}. There are 17 parameters (only 5 parameters
characterize the GM models \cite{Glendenning:1991es}) which are fixed
by means of a global fit on nuclear matter and finite nuclei's
properties. For our discussion the crucial quantity is the symmetry
energy and its derivative with respect to the density: here we adopt
the parametrization called SFHo for which $S=32$ MeV (very close to
the GM3 value) and $L=47$ MeV. We have included in the lagrangian hyperons (assuming SU(6) symmetry) and $\Delta$
resonances (assuming $x_{\sigma\Delta}=x_{\rho\Delta}=1$ and three different values for $x_{\omega\Delta}$).
Results for the particles' fractions
as functions of the baryon density in $\beta$-stable matter are displayed
in Fig. \ref{particles}. In the upper panel, we have included only
hyperons: the $\Lambda$ and the $\Xi^-$ appear at a density of about
$0.5$ fm$^{-3}$ and then the $\Xi^0$ at a density of about $1.1$
fm$^{-3}$. The $\Sigma$ hyperons are strongly suppressed because of
their repulsive optical potential and are basically irrelevant for the
structure of neutron stars. In the lower panel we include also the
$\Delta$ isobars. In agreement with what found from the previous analysis, for values of
$L$ smaller than about $65$ MeV the $\Delta$s also appear at densities
relevant for neutron stars and actually, in the SFHo model, they
appear even before the hyperons with the $\Delta^-$ formed at a
density of about $0.4$ fm$^{-3}$. The appearance of these particles
delays the appearance of hyperons: the threshold for the $\Xi^-$ is
shifted to higher densities by about $0.15$ fm$^{-3}$. The $\Lambda$
is also slightly shifted to higher densities in agreement with the
results of Fig. \ref{soglie}. It is important to remark that, within the SFHo model, even
using $x_{\omega\Delta}=1.1$, the $\Delta^-$ appear before hyperons.

Let us now discuss the uncertainties on the couplings between
$\Delta$s and mesons. Qualitatively, it has been possible to establish
that the $\Delta$s inside a nucleus feel an attractive
potential. There are several purely theoretical studies on the
properties of the isobars in the nuclear medium: for instance, in
Ref.\cite{Jin:1994vw}, from QCD sum rules, it has been found that
$x_{\omega\Delta}$ is significantly smaller than 1. In the many body
analysis of Ref.\cite{Oset:1987re}, the real part of the $\Delta$
self-energy has been evaluated to be about $-30$ MeV at $n_B=0.75
n_0$. Notice that this self energy is relative to the one of the
nucleon and the total potential felt by the $\Delta$ is the sum of its
self energy and of the nucleon potential, a number of the order of
$-80$ MeV \footnote{E. Oset, private communications}.  Also
phenomenological analysis have been performed of data from
electron-nucleus
\cite{Koch:1985qz,Wehrberger:1989cd,O'Connell:1990zg}, photoabsorption
\cite{Alberico:1994sx} and pion-nucleus scattering
\cite{Horikawa:1980cv,Nakamura:2009iq}.  When discussing pion
scattering data, a value for the real part of the $\Delta-$nucleus
potential of $-30$ MeV is extracted \cite{Horikawa:1980cv}. Since
pions interact mainly with the nuclear surface, larger values are
expected for the binding at $n_0$. More recently a global analysis of
pion-nucleus scattering and of pion photo-production has been
performed in Ref.~\cite{Nakamura:2009iq} where the experimental data
are correctly described by assuming a $\Delta$ potential equal to the
nucleon potential. From the data analysis of electron-nucleus
scattering, either density or momentum dependent potentials have been
deduced. In Ref.\cite{Koch:1985qz} the binding potential is
parameterized as $-75 \,n_B(r)/n_0$ MeV. In
Ref.\cite{O'Connell:1990zg} they obtain an optical potential which, at
a momentum of about $400$ MeV (quite typical for electron scattering),
gives a binding in agreement with the one of \cite{Koch:1985qz}.
Electromagnetic excitations of the $\Delta$ baryon have been also
analyzed within a relativistic quantum hadrodynamics scheme with the result that
$0\lesssim x_{\sigma\Delta}-x_{\omega\Delta}\lesssim 0.2$
\cite{Wehrberger:1989cd}. The conclusion one can draw from all these
analysis is that the potential of the  $\Delta$ falls within the
range -$30$ MeV $+ V_N \lesssim V_{\Delta} \lesssim V_N$ where $V_N$
is the nucleon potential.

In the relativistic mean field model \cite{Glendenning:1991es} the
potential of the $\Delta$ (which coincides with the binding energy of
the lowest $\Delta$ level) is given by: $V_{\Delta}=x_{\omega\Delta}
g_{\omega n}\omega-x_{\sigma\Delta} g_{\sigma n} \sigma$ where the
mean fields are calculated at $n_0$. By fixing a value for
$V_{\Delta}$ a relation between $x_{\sigma\Delta}$ and
$x_{\omega\Delta}$ is obtained, shown in Fig.~\ref{be} together with
the experimental constraints on $x_{\sigma\Delta}-x_{\omega\Delta}$.
New analysis, and possibly new experiments, aiming at a better
determinations of these couplings would be extremely important.
Notice also that no information is available for $x_{\rho\Delta}$
which in principle could be extracted by analyzing scattering on
neutron rich nuclei.

\begin{figure}[tb]
\vskip 0cm
\begin{centering}
\epsfig{file=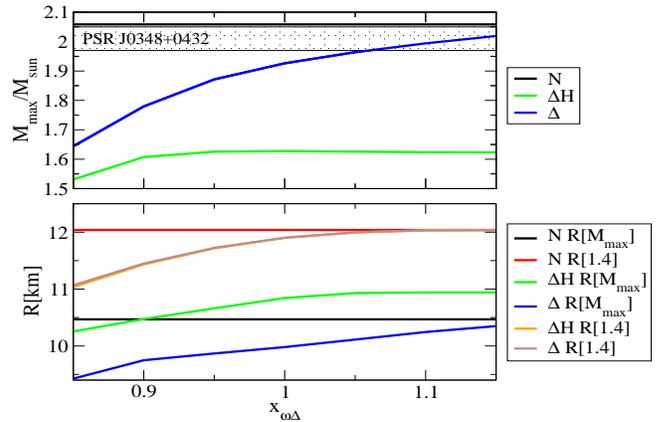,height=8.5cm,width=5.5cm,angle=-90}
\caption{(Color online) Properties of hadronic stars (with and without hyperons)
as functions of $x_{\omega\Delta}$: the maximum mass is displayed in the upper panel while
the radii of the $1.4 M_{\odot}$ stellar configurations and the radii of the maximum mass configurations are displayed in the lower panel.
The labels $N$, $\Delta$ and $\Delta H$ in the legend stand for purely nucleonic stars, for hadronic stars with only $\Delta$s and for hadronic stars
in which $\Delta$s and hyperons are present. The radii of the $1.4 M_{\odot}$ $\Delta H$ hadronic stars coincide with the ones of $\Delta$ hadronic stars because
hyperons do not appear in those stellar configurations. Since the maximum mass of the $\Delta H$ configuration is smaller than the one of the $\Delta$ configurations, the corresponding
radius is larger (see Fig.1 of Ref.~\cite{Drago:2013fsa}).}
\label{mr1}
\end{centering}
\end{figure}

Let us now analyze the effect of including $\Delta$s on the structure of
neutron stars. We calculate the equation of state of $\beta$-stable
matter by use of the SFHo model for different values of
$x_{\omega\Delta}$ at fixed values of
$x_{\sigma\Delta}=x_{\rho\Delta}=1$ (similar results are found by
varying $x_{\sigma\Delta}$). From the upper panel of Fig.~\ref{mr1} one
can notice that the inclusion of the $\Delta$ dramatically reduces the
maximum mass: if $x_{\omega\Delta} \lesssim 1$ as indicated by the experimental
data, the maximum mass does not satisfy the $2 M_{\odot}$ limit
\cite{Antoniadis:2013pzd}.
Concerning the radii, we notice that if only $\Delta$
resonances are included the maximum mass configurations are very
compact, with a radius $R \lesssim 10.5$ km. Concerning hyperons, we
have not taken into account possible mechanisms making the equation of
state stiffer at high densities such as the inclusion of the $\phi$
meson \cite{Weissenborn:2011ut,Weissenborn:2011kb}. The reason is
that in this work we are interested in showing that already the
appearance of $\Delta$s can lead to a problem with astrophysical
measurements. The implementation of additional repulsion between
hyperons would shift the green curves towards the blue ones which
correspond to the case in which hyperons are not present at all.

To summarize: we have shown that recent constraints on the value of
the density derivative of the symmetry energy indicate, indirectly, an
early appearance of $\Delta$ isobars in $\beta$-stable matter, at a
density of the order of $2\div 3\, n_0$. These degrees of freedom are
therefore necessary ingredients of the equation of state of neutron
star matter. In turn their appearance modify the composition and the
structure of hadronic stars. In particular the effect on the maximum
mass is rather dramatic. If their potential is of the order of the
one indicated by analysis of laboratory data on pion and
electron-nucleus scattering and on photoabsorption nuclear reactions
then the maximum mass is below the $2 M_{\odot}$ limit and we are
facing a $\Delta$-isobar puzzle in the physics of neutron stars.

\vskip 0.5cm
We thank M. B. Barbaro and E. Oset for very useful discussions.
G.P. acknowledges financial support from the Italian Ministry of Research through the
program \textquotedblleft Rita Levi Montalcini\textquotedblright.


\end{document}